\documentclass{ifcolog-c}

\title{Recursion does not always help}
\titlethanks{}
\addauthor[gdp@inf.ed.ac.uk]{Gordon Plotkin}{University of Edinburgh}

\usepackage{amsmath,amsthm,amssymb}
\usepackage{color}

\authorrunning{Plotkin}
\titlerunning{Recursion does not always help }

\date{January 2022}
\newcommand{\Y}{\mathrm{Y}}

\newcommand{\eqdef}{=_{\small \mathrm{def}}}

\newcommand{\cut}[1]{}

\newtheorem{lemma}{Lemma}
\newtheorem{theorem}{Theorem}
 \newtheorem{corollary}{Corollary}
 
\begin{document}
\maketitle
\begin{abstract} We show that 
adding recursion does not increase the  total  functions 
definable in the typed $\lambda\beta\eta$-calculus  or 
the partial functions 
definable in the $\lambda\Omega$-calculus.  As a consequence, adding recursion does not increase the class of partial or total definable functions on free algebras  and so, in particular, on the natural numbers.

\end{abstract}
\section{Introduction}
 
 As is well known, using  Church  numerals as ``codes'' for natural numbers, all partial recursive functions can be defined in  the untyped 
$\lambda$-calculus. The definitions make use of recursion. % via the $\Y$-combinator. 
If we switch to a  typed framework, the situation changes drastically: in the typed $\lambda\beta\eta$-calculus (or the $\lambda\beta$-calculus) only the extended polynomials can be defined if we use Church numerals of type
 $(o \rightarrow o) \rightarrow (o \rightarrow o)$ to represent the set of natural numbers, see~\cite{HS75,Sta79}. 
 
 Further functions can be defined if, more generally, one  uses representing types of the form  $(\sigma \rightarrow \sigma) \rightarrow (\sigma \rightarrow \sigma)$.  Fortune et al~\cite{FLO83} showed that this is possible non-uniformly, that is, if one uses different types 
  for  arguments or  results. For example, predecessor can be defined non-uniformly, but, as shown by  Zakrzewski~\cite{Zak07},   cannot be defined uniformly.  Zakrzewski further showed---contrary to the then common belief---that there are uniform examples. In particular he showed that, for $l \geq 2$, the function 
  \[f(m,n_0,\ldots,n_{l-1}) = n_{m \,\mathrm{mod}\, l}\]
is uniformly definable, as is, for $l \geq 1$, the characteristic function of the predicate $\leq l$. He further conjectured that adding these  to the schema for the extended polynomials characterises all the uniformly definable functions.
 But, in any case, one cannot so represent all total recursive functions. 
 It is therefore natural to ask which partial numerical functions can be defined if one adds recursion at all types, i.e., if one uses the $\lambda\Y$-calculus~\cite{Sta04}. 
 
 More generally than numerical functions, free algebras can be represented in the $\lambda\beta\eta$-calculus and then functions on  free algebras can be defined, see~\cite[p.\ 38]{BDS13}. So one can again ask whether adding recursion increases the class of definable functions. More generally still, given representations of non-empty sets $X_i \; (i =1,\ldots, k)$, and $X$ and corresponding coding functions, one can ask if adding recursion changes the set of  partial  definable functions $f: X_1\times \ldots \times X_k \rightharpoonup X$.
 
 Perhaps surprisingly, it turns out (Theorem~\ref{myresult}) that %, under mild assumptions on the representing types, 
 no more total functions can be defined  in the $\lambda\Y$-calculus than in the $\lambda\beta\eta$-calculus, and, further, %without any assumptions, 
 that no more partial functions can be defined  in the $\lambda\Y$-calculus than in the $\lambda\Omega$-calculus (the $\lambda\beta\eta$-calculus extended with a ground type constant $\Omega$ for ``undefined''). So, in particular, 
 %(Corollary~\ref{mycorollary}) 
 adding recursion does not increase the available definable partial or total functions on free algebras. The reason is that it is not possible to make use of unbounded recursion depth in the $\lambda\Y$-calculus.

 \cut{
 Perhaps surprisingly, it turns out (Theorem~\ref{myresult}) that no more total numerical functions can be defined than can be defined in the $\lambda\beta\eta$-calculus, and that no more partial numerical functions can be defined than can be defined in the $\lambda\Omega$-calculus---the $\lambda\beta\eta$-calculus extended with a ground type constant $\Omega$. The reason is that it is not possible to make use of unbounded recursion depth in the $\lambda\Y$-calculus.   
 }

 The main tool we use to prove our results is due to Werner Damm~\cite{Dam82}.  It is a kind of ``higher-order flow analysis'' using a finite domain. It can also be used to prove results of Statman, that the termination of closed 
 $\lambda\Y$-terms is decidable, as is whether they have a head normal form.

 It is worth remarking that, in contrast, in the second-order $\lambda$-calculus where the type of the numerals
 is $\forall X. (X\rightarrow X) \rightarrow (X \rightarrow X)$, all functions provably total in second-order Peano arithmetic, and so all primitive recursive functions, are 
 definable, see~\cite{Sta81,GTL89}.  
 If recursion is added, all total recursive functions become definable.

 This paper is written in honour of Jonathan Seldin on the occasion of his 80th birthday.
\section{Definable  functions }

The typed $\lambda\beta\eta$-calculus is as in~\cite{Bar82}, say, with types built up from a single ground type $o$, and, following Church, with variables $x^\sigma$  carrying their own types. 
We may omit the types when they can be understood from the context. We write types of the form  $\sigma_1 \rightarrow \ldots \rightarrow \sigma_n \rightarrow o$ as $(\sigma_1, \ldots, \sigma_n)$ and say they are $n$-ary. We write $\Lambda_\sigma$ for the set of closed terms of type $\sigma$. 

By an \emph{extension} $\lambda^+$ of the $\lambda\beta\eta$-calculus, we mean a calculus whose terms are $\lambda$-terms with additional constants, and with a type-respecting conversion relation between those terms, written:
\[\vdash_{\lambda^+} M = N\]
that is an equivalence relation closed under $\lambda$-abstraction, application, and substitution, and which contains 
$\lambda\beta\eta$-conversion. 

Such a calculus $\lambda^+_2$ is an \emph{extension} of another such calculus $\lambda^+_1$ if the constants of $\lambda^+_1$ are also constants of 
$\lambda^+_2$ and, for any $\lambda^+_1$- terms $M$ and $N$, we have:
\[\vdash_{\lambda^+_1} M = N \implies \vdash_{\lambda^+_2} M = N\]
The extension is \emph{conservative} if, for any two such terms, the reverse implication also holds.

 A \emph{coding} function for a set $X$ is simply a function from $X$ to some $\Lambda_\sigma$ ($\sigma$ is the representing type). 
 %We say that $\gamma: X \rightarrow \Lambda_\sigma$ is a \emph{coding} function if it is \emph{$\lambda$-injective}, by which we %mean:
%
%\[\vdash_{\lambda\beta\eta} \gamma(x) = \gamma(y) \implies x = y\]
%
Let $\lambda^+$ be an extension of the $\lambda\beta\eta$-calculus. Then, for nonempty sets $X_i$ ($i = 1,\ldots,k$) and $X$, given coding functions $\gamma_i: X_i \rightarrow \Lambda_{\sigma_i}$ ($i = 1,\ldots,k$), and $\gamma: X \rightarrow \Lambda_\sigma$, a closed term 
$F: \sigma_1 \rightarrow \ldots \rightarrow \sigma_k \rightarrow \sigma$  %(where $k \geq 0$)  
is said to \emph{$(\gamma_1, \ldots, \gamma_k, \gamma)$-define} a  partial  %partial 
function %$f: \mathbb{N}^k \rightarrow \mathbb{N}$ %
$f: X_1 \times \ldots \times X_k \rightharpoonup X$ in $\lambda^+$
 if  the following two conditions hold:
%
%\[f(x_1,\ldots,x_k) \simeq x \iff \vdash_{\lambda^+} F\gamma(x_1) \ldots \gamma(x_k) = \gamma(x)\]
\[f(x_1,\ldots,x_k) = x \implies \vdash_{\lambda^+} F\gamma(x_1) \ldots \gamma(x_k) = \gamma(x)\qquad (\mbox{Graph})\]
and
\[f(x_1,\ldots,x_k)\uparrow \implies \not\exists N\in \Lambda_\sigma.\, \vdash_{\lambda^+} F\gamma(x_1) \ldots \gamma(x_k) = N \qquad (\mbox{Undef})\]
 If  $f$ is total this amounts to
\[\vdash_{\lambda^+} F\gamma(x_1) \ldots \gamma(x_k) = \gamma(f(x_1,\ldots,x_k))\]
 If $\gamma_1 = \ldots = \gamma_k = \gamma$, we say that $F$ 
\emph{$\gamma$-defines} $f$ and that $f$ is \emph{uniformly} definable. Note that, with this definition, only total functions can be defined in $\lambda\beta\eta$.
%\footnote{We use Kleene equality here, where one writes $e \simeq e'$ for two mathematical expressions to mean 
%that either they are both defined and equal or else they are both undefined.}. 
%

Algebraic datatypes provide examples, as explained in~\cite[p.\ 38]{BDS13}. There free algebras 
$\mathcal{A}_\Sigma$ over given signatures $\Sigma$ are considered, and \emph{standard representation} types 
$\rho_\Sigma$  and coding functions $\gamma_\Sigma: \mathcal{A}_\Sigma \rightarrow \rho_\Sigma$ are given.  The nontrivial case (the one where the free algebra is nonempty) is where the signature includes at least one constant; in that case the representation type  is inhabited. In the case of the natural numbers, the signature $\Sigma_{{\small \mathrm{Nat}}}$ is that of one unary function and one constant, the standard representation type is 
$(o \rightarrow o) \rightarrow (o \rightarrow o)$, and the Church numerals provide the standard coding function:
\[\gamma_{\Sigma_{{\small \mathrm{Nat}}}}(m) \eqdef \lambda f^{o\rightarrow o}.\lambda x^o.f^m(x)\]

The free algebra functions and combinations of them using certain representable discriminators are shown standardly definable. Zaionc~\cite{Zai91} has given an exact characterisation of the standardly definable functions in terms of a certain limited recursion scheme, as well as more explicit characterisations for the cases of trees and words (this last generalising that for  the natural numbers).

One can proceed somewhat more generally. One uses instances $\rho_{\Sigma,\sigma} \eqdef \rho_\Sigma[\sigma/o]$ of the standard representation type, and one then obtains  corresponding  coding functions $\gamma_{\Sigma,\sigma}: \mathbb{A}_\Sigma \rightarrow \rho_{\Sigma,\sigma}$.
We refer to $(\gamma_{\Sigma, \sigma_1}, \ldots, \gamma_{\Sigma, \sigma_k}, \gamma_{\Sigma,\sigma})$-definability (respectively $\gamma_{\Sigma,\sigma}$-definability) as 
$(\sigma_1, \ldots, \sigma_k,\sigma)$-definability (respectively $\sigma$-definability).
Taking instances, we note that $o$-definable functions are  $\sigma$-definable, for any $\sigma$.
%for any $\beta$. 

\cut{ Church numerals provide a common example. The type $\omega_\beta$ is
\[(\beta \rightarrow \beta) \rightarrow (\beta \rightarrow \beta) \]
The numeral for $m \in \mathbb{N}$ at $\beta$ is
\[\underline{m}_\beta \eqdef \lambda f^{\beta \rightarrow \beta}.\lambda x^\beta.f^m(x)\]
and we have a coding function $\gamma_\beta: \mathbb{N} \rightarrow \omega_\beta$ where $\gamma_\beta(m) = \underline{m}_\alpha$. We refer to $(\gamma_{\beta_1}, \ldots, \gamma_{\beta_k}, \gamma_\beta)$-definability (respectively $\gamma_\beta$-definability) as 
$(\beta_1, \ldots, \beta_k,\beta)$-definability (respectively $\beta$-definability).
}

%A closed term $F: \omega_{\sigma_1} \rightarrow \ldots \rightarrow \omega_{\sigma_k} \rightarrow \omega_\alpha$ in 
%an extension $\lambda^+$ of the $\lambda\beta\eta$-calculus %(where $k \geq 0$)  
%is said to \emph{$(\sigma_1, \ldots, \sigman, \alpha)$-define} a  partial  numerical %partial 
%function %$f: \mathbb{N}^k \rightarrow \mathbb{N}$ %
%$f: \mathbb{N}^k \rightharpoonup \mathbb{N}$ 
% if, and only if,   we have:
%%
%\[f(m_1,\ldots,m_k) \simeq m \iff \vdash_{\lambda^+} F\underline{m_1}_{\sigma_1} \ldots \underline{m_k}_{\sigma_k} = 
%\underline{m}_\alpha\]
%%
%for all $m_1,\ldots,m_k,m \in \mathbb{N}$. If $\sigma_1 = \ldots = \sigma_k = \alpha$, we say $F$ 
%\emph{$\alpha$-defines} $f$ and that $f$ is \emph{uniformly} definable. (We use Kleene equality here, where one 
%writes $e \simeq e'$ for two mathematical expressions to mean that either they are both defined and equal or else 
%they are both undefined.)

\newcommand{\tr}{\mathrm{tr}}
Returning to general considerations, let $\lambda^+_1$ and $\lambda^+_2$ be extensions of the $\lambda\beta\eta$-calculus. Then $\lambda^+_1$  and 
$\lambda^+_2$   are \emph{equipotent} for  $(\gamma_1, \ldots, \gamma_k, \gamma)$-definable (total)  functions, if a (total) function is $(\gamma_1, \ldots, \gamma_k, \gamma)$-definable in $\lambda^+_1$ if, and only if,  it is  in $\lambda^+_2$. 

The next lemma presents a strategy for showing that all functions definable in one extension of the $\lambda\beta\eta$-calculus are definable in another. It gives requirements on a translation from one calculus to another sufficient to ensure that every function definable in the first is also definable in the second.

\newpage 

\begin{lemma} \label{general} Let $\lambda^+_1$ and $\lambda^+_2$ be extensions of the $\lambda\beta\eta$-calculus, and let $\tr$ be a type and application respecting map from  closed $\lambda^+_1$-terms to closed $\lambda^+_2$-terms which acts as the identity on closed $\lambda$-terms. Suppose further that, for every closed $\lambda^+_1$-term $M$ and closed $\lambda$-term $N$ of the same type we have:
\[\vdash_{\lambda^+_1} M = N \iff  \vdash_{\lambda^+_2}\tr(M) = N\qquad (*)\]
Then every function $(\gamma_1, \ldots, \gamma_k, \gamma)$-definable  in $\lambda^+_1$ is $(\gamma_1, \ldots, \gamma_k, \gamma)$-definable in $\lambda^+_2$. Further, in the case such a function is total, only the implication from left to right in $(*)$ is needed.
\end{lemma}
\begin{proof} Let  $\gamma_i: X_i \rightarrow \Lambda_{\sigma_i}$ ($i = 1,\ldots,k$), and $\gamma: X \rightarrow \Lambda_\sigma$ be coding functions, and suppose that the term $F: \sigma_1 \rightarrow \ldots \rightarrow \sigma_k \rightarrow \sigma$ 
$(\gamma_1, \ldots, \gamma_k, \gamma)$-defines
%$f: \mathbb{N}^k \rightarrow \mathbb{N}$ %
$f: X_1 \times \ldots \times X_k \rightharpoonup X$  in $\lambda^+_1$.

Using the assumptions on $\tr$  and $(*)$ we have the following implications for the Graph condition.
\[\begin{array}{lcl}f(x_1,\ldots,x_k) = x &\! \implies\!&  \vdash_{\lambda_1}F\gamma_1(x_1)\ldots\gamma_k(x_k) = \gamma(x)\\
                                                             &  \! \implies\! &  \vdash_{\lambda_2}\tr(F\gamma_1(x_1)\ldots\gamma_k(x_k)) = \gamma(x)\\
                                                            &  \! \implies\!  &\vdash_{\lambda_2}\tr(F)\gamma_1(x_1)\ldots\gamma_k(x_k) = \gamma(x)\\
\end{array}\]
and, for any $N\in \Lambda_\sigma$,  the following implications for the Undef condition:
\[\begin{array}{lcl}
\vdash_{\lambda_2}\tr(F)\gamma_1(x_1)\ldots\gamma_k(x_k) = N \! 
& \!\!  \! \implies\!\!\!  &\!  \vdash_{\lambda_2}\tr(F\gamma_1(x_1)\ldots\gamma_k(x_k)) = N \\
\! & \! \! \! \implies\! \!\!  &\!  \vdash_{\lambda_1}F\gamma_1(x_1)\ldots\gamma_k(x_k) = N \\
                                                            &  \! \implies\!  &f(x_1,\ldots,x_k)\downarrow\\
                                                            \end{array}\]
\end{proof} 
Note that in the case $\lambda^+_2$ is a conservative extension of $\lambda^+_1$, one can take the translation $\tr$ to be the identity (with only an extension being needed in the case of total functions).

%Let $\lambda^+$ and $\lambda^{++}$ be extensions of the $\lambda\beta\eta$-calculus, with $\lambda^{++}$ 
%extending $\lambda^{+}$. Then $\lambda^{++}$ is \emph{conservative} over $\lambda^{+}$   for  $(\sigma_1, \ldots, 
%\sigman, \alpha)$-definable (total) numerical functions, if every such function that is $(\sigma_1, \ldots, \sigman, %\alpha)$-definable in $\lambda^+$ is $(\sigma_1, \ldots, \sigman, \alpha)$-definable in $\lambda^{++}$. 

\section{The $\lambda\Omega^+$-calculus}
The $\lambda\Omega^+$-calculus is the $\lambda\beta\eta$-calculus extended with  constants $\Omega_\sigma:\sigma$ for every type $\sigma$, and the conversion relation generated by  $\beta$- and $\eta$-conversion for all terms. With $\beta$-reduction and reverse $\eta$-reduction, one has Church-Rosser and  long $\beta\eta$-normal forms, just as in the $\lambda\beta\eta$-calculus.
Long $\beta\eta$-normal forms containing no  $\Omega_\sigma$  are called \emph{proper}, all others are called \emph{improper}.  
The $\lambda\Omega$-calculus is the subcalculus with just the constant $\Omega_o$ (written~$\Omega$). 
The $\lambda\Omega^+$-calculus is a useful intermediary between the $\lambda\Y$-calculus and the $\lambda\Omega$-calculus: the $\Omega_\sigma$ act as variables of type $\sigma$ which can be substituted for to link with those calculi.

The $\lambda\Omega^+$-calculus is a conservative extension of the $\lambda\Omega$-calculus, and, in turn, the $\lambda\Omega$-calculus is a conservative extension of the $\lambda\beta\eta$-calculus. By the remark after 
Lemma ~\ref{general} it follows that every function definable in the $\lambda\beta\eta$-calculus is definable in the $\lambda\Omega$-calculus,  and that every  function definable in the $\lambda\Omega$-calculus is definable in the $\lambda\Omega^+$-calculus (all with the same coding scheme).

\begin{lemma}  \label{downOmega}
The $\lambda\beta\eta$  and 
$\lambda\Omega$  calculi  are \emph{equipotent} for   $(\gamma_1, \ldots, \gamma_k, \gamma)$-definable  total functions. 
\end{lemma}
\begin{proof}   We have already seen that every total function $(\gamma_1, \ldots, \gamma_k, \gamma)$-definable in $\lambda\beta\eta$ is $(\gamma_1, \ldots, \gamma_k, \gamma)$-definable in $\lambda\Omega$.  Conversely, suppose that $F:\sigma_1\rightarrow \ldots \rightarrow \sigma_k \rightarrow \sigma$ $(\gamma_1, \ldots, \gamma_k, \gamma)$-defines a total function $f$ in the $\lambda\Omega$-calculus. 
The long $\beta\eta$-normal form of $F$ has the form
%
%\[\lambda n_1\ldots \lambda n_1 \lambda f^{\sigma \rightarrow \sigma} \lambda a^\sigma \lambda b^{\tau_1}\lambda b_1^{\tau_1} \ldots \lambda b_l^{\tau_l} M[\Omega] \]
\[F[\Omega] = \lambda x_1^{\sigma_1}\ldots  x_k^{\sigma_k}. \lambda y_1^{\tau_1} \ldots y_l^{\tau_l}. M[\Omega]  \]
with $M[\Omega]: o$,  where $\sigma = (\tau_1,\ldots,\tau_l)$. Choosing $a \in X$ (recall that $X$ is non-empty) we obtain a closed term $A \eqdef \gamma(a):\sigma$

%Choose a closed term $M: \sigma_1 \rightarrow \ldots  \rightarrow \sigma_k \rightarrow \sigma$.  
Then
\[ F[Ay_1^{\tau_1}\ldots y_l^{\tau_l}]  = \lambda x_1^{\sigma_1}\ldots  x_k^{\sigma_k}. \lambda y_1^{\tau_1} \ldots y_l^{\tau_l}. M[Ay_1^{\tau_1}\ldots y_l^{\tau_l}] \]
is a $\lambda$-term defining the same function, as the function is total.
and  
\[\vdash_{\lambda\Omega} F[\Omega]  \gamma_1(a_1)\ldots\gamma_1(a_k) = \gamma(f(a_1,\ldots,a_k))\]
implies
 \[\vdash_{\lambda\beta\eta} F[Ay_1^{\tau_1}\ldots y_l^{\tau_l}]  \gamma_1(a_1)\ldots\gamma_1(a_k) = \gamma(f(a_1,\ldots,a_k))\]
\end{proof}

There is a natural translation from  $\lambda\Omega^+$ to $\lambda\Omega$. For $n$-ary  $\sigma$, set
$\overline{\Omega}_\sigma = \lambda x_1\ldots x_n. \Omega$
. Then, for any $\lambda\Omega^+$-term $M$, let $\overline{M}$ be the $\lambda\Omega$-term obtained from $M$ by replacing all the $\Omega_\sigma$'s in $M$ by $\overline{\Omega}_\sigma$'s. 
%
%The following lemma provides a syntactic link between $\lambda\Omega^+$ and $\lambda\Omega$. %We omit the evident proof.
\begin{lemma} \label{downsynt}For any $\lambda\Omega^+$-term $M$, and any $\lambda$-term $N$ of the same type we have:
\[\vdash_{\lambda\Omega^+} M = N \implies \vdash_{\lambda\Omega} \overline{M} = N\]
\end{lemma}
\begin{proof} As the $\Omega_\sigma$ act as variables, one can replace the $\Omega_\sigma$ in $M$ and $N$ by the corresponding 
$\overline{\Omega}_\sigma$. As this changes $M$ to $\overline{M}$ and leaves $N$ alone, the result follows.
\end{proof}
It would be interesting to have a syntactic proof of the converse of this lemma;  a flow-analyis proof is given below.

\section{The $\lambda\Y$-calculus}

The $\lambda\Y$-calculus~\cite{Sta04} is the $\lambda\beta\eta$-calculus extended with  recursion operators, i.e., constants
\[\Y_\sigma: (\sigma \rightarrow \sigma) \rightarrow \sigma\]
and conversions $\Y_\sigma F = F(\Y_\sigma f)$, for $F:\sigma \rightarrow \sigma$. With reduction rules $\beta$, $\eta$, and 
$\Y_\sigma  \rightarrow \lambda f. f(\Y_\sigma f)$, it is Church-Rosser (see~\cite{Sta04} and~\cite[p.\ 314]{BDS13}).

There is a natural translation from  $\lambda\Omega^+$ to $\lambda\Y$. Set
$\widehat{\Omega}_\sigma = \Y_\sigma(\lambda x.\,x)$. 
Then, for any $\lambda\Omega^+$-term $M$, let $\widehat{M}$ be the $\lambda\Y$-term obtained from $M$ by replacing all the $\Omega_\sigma$'s in $M$ by $\widehat{\Omega}_\sigma$'s. We evidently have:
\begin{lemma} \label{hat}For any $\lambda\Omega^+$-term $M$, and any $\lambda$-term $N$ of the same type we have:
\[\vdash_{\lambda\Omega^+} M = N \implies \vdash_{\lambda\Y} \widehat{M} = N\]
\end{lemma}
As in the case of Lemma~\ref{downsynt}, the converse will be obtained by flow-analysis.

In the other direction, we work with particular approximations to $\Y$ in the next section. We first consider such approximations in general.
Define $\lambda\Omega^+$-terms $\widetilde{\Y}^{(n)}_\sigma: (\sigma \rightarrow \sigma) \rightarrow \sigma$ 
by $\widetilde{\Y}^{(n)}_\sigma = \lambda f.f^n(\Omega_{\sigma\rightarrow \sigma}f)$ and 
$\lambda\Y$-terms $\Y^{(n)}_\sigma: (\sigma \rightarrow \sigma) \rightarrow \sigma$ by $\Y^{(n)}_\sigma = \lambda f.f^n(\Y_\sigma(f))$.
Then, for any $\lambda\Y$-term $M[\Y_{\sigma_1},\ldots, \Y_{\sigma_k}]$ and any $n_1,\ldots,n_k$, let  $M^{(n_1,\ldots,n_k)}$ be the $\lambda\Omega^+$-term $M[\widetilde{\Y}_{\sigma_1}^{(n_1)},\ldots, \widetilde{\Y}_{\sigma_n}^{(n_k)}]$.

\begin{lemma} \label{upsynt}For any $\lambda\Y$-term $M$, and any $\lambda$-term $N$ of the same type we have:
\[\vdash_{\lambda\Omega^+} M^{(n_1,\ldots,n_k)} = N \implies \vdash_{\lambda\Y} M = N\]
\end{lemma}
\begin{proof} We have  $\vdash_{\lambda\Omega^+} M[\widetilde{\Y}_{\sigma_1}^{(n_1)},\ldots, \widetilde{\Y}_{\sigma_k}^{(n_k)}] = N$. As the $\Omega_{\sigma \rightarrow \sigma}$ act as variables, and do not occur in $N$,  $\vdash_{\lambda\Y} M[\Y_{\sigma_1}^{(n_1)},\ldots, \Y_{\sigma_k}^{(n_k)}] = N$ (replacing $\Omega_{\sigma \rightarrow \sigma}$'s by $\Y_\sigma$'s). The conclusion follows, as, for any $\sigma$ and $n$,  we have $\vdash_{\lambda\Y} \Y_{\sigma}^{(n)} = \Y_\sigma$.

\end{proof}
We remark that if a version of this lemma for the $\lambda\Omega$-calculus were available, the 
$\lambda\Omega^+$-calculus would not be needed.
\section{Semantics}
\renewcommand{\O}{\mathcal{O}}
\newcommand{\sem}[1]{[\!|#1|\!]}

We work over the simple type hierarchy $\O_\sigma$ of continuous functions starting with  Sierpi{\'n}ski space: $\O_o = \mathbb{O}$ (i.e., $\{\perp,\top\}$, with $\perp \leq \top$). As these domains are all finite, this is also the hierarchy of monotone functions. This gives an interpretation of the typed $\lambda\beta\eta$-calculus in a standard way.
We write $\O\sem{M}(\rho)$ for the interpretation of a term $M$ in environment $\rho$, and generally omit the $\rho$ when $M$ is closed.
 The interpretation is extended to the $\lambda\Omega^+$-calculus by taking  $\O\sem{\Omega_\sigma}$ to be $\perp_{\O_\sigma}$ and to the $\lambda\Y$-calculus by taking $\O\sem{\Y_\sigma}$ to be the least fixed point operator 
 $f \mapsto \bigvee_n f^n(\perp_{\O_\sigma})$. 
 
 We have $\mathcal{O}\sem{\Omega_\sigma} = \mathcal{O}\sem{\overline{\Omega}_\sigma} = \mathcal{O}\sem{\widehat{\Omega}_\sigma} $. So, for any closed $\lambda\Omega^+$-term $M$, we have: $\mathcal{O}\sem{M} = \mathcal{O}\sem{\overline{M}} = \mathcal{O}\sem{\widehat{M}}$.
 Next, taking $h(\sigma)$ to be the height of the longest ascending chain in $\O_\sigma$, the least-fixed-point operator is the same as the truncated operator $f \mapsto f^{h(\sigma)}(\perp_{\O_\sigma})$, and that is precisely $\mathcal{O}\sem{\widetilde{\Y}^{(h(\sigma))}_\sigma}$. 
 
 We can now define the translation from $\lambda\Y$ to $\lambda\Omega^+$. For any closed $\lambda\Y$-term 
 $M[Y_{\sigma_1}\ldots,Y_{\sigma_k}]$, 
 % $M$ containing the operators $Y_{\sigma_1}\ldots,Y_{\sigma_n}$, 
 set $\widetilde{M} = M^{((h(\sigma_1),\ldots,(h(\sigma_k))}$. 
 %let $\widetilde{M}$ be the closed $\lambda\Omega^+$-term obtained from $M$ by replacing the operators by $
 %\widetilde{\Y}^{(h(\sigma_1))},\ldots,\widetilde{\Y}^{(h(\sigma_n))}$ respectively. 
 Note that we  then have:  $\mathcal{O}\sem{M} = \mathcal{O}\sem{\widetilde{M}} $. 

As we shall see, $\widetilde{M}$ encodes the maximum recursion depth available to the recursion operators in $M$. To show this, as mentioned in the introduction,  we employ  Werner Damm's higher-order flow analysis over a finite model. The analysis is carried out using certain ``test functions''   $t_\sigma: \O_\sigma \rightarrow \mathbb{O}$. As shown in Lemma~\ref{flow} below, these test functions distinguish proper from improper normal forms.

The $t_\sigma$
  are defined mutually inductively with $s_\sigma \in \O_\sigma$ by setting:
\[t_\sigma(f) = fs_{\sigma_1}\ldots s_{\sigma_n}
\qquad  \qquad s_\sigma f_1\ldots f_n = \bigwedge_i t_{\sigma_i}(f_i)\]
for $\sigma = (\sigma_1,\ldots, \sigma_n)$.
Note that, in particular, $t_o(x) = x$ and $s_o = \top$; we further have 
$t_{\sigma \rightarrow \tau}f = t_\tau(f s_\sigma)$.
%

%The next lemma shows that these tests  distinguish between proper and improper normal forms.
\begin{lemma} \label{flow}
Let $M: \sigma$ be a  long $\beta\eta$-normal form in $\lambda\Omega^+$.
Then:
\[t_\sigma(\O\sem{M}) = \top \iff \mbox{M is proper}\]
\begin{proof}
Let $\sigma$ be $(\sigma_1,\ldots,\sigma_n)$. We proceed by induction. Suppose that $M$ is proper. Then $M$ has the form
\[\lambda f_1\ldots f_n. f_{i_0} M_1\ldots M_k\]
where 
$\sigma_{i_0} = (\tau_1,\ldots, \tau_k)$
and the $N_j \eqdef  \lambda f_1\ldots f_n. M_j$ are strictly smaller closed long $\beta\eta$-normal forms. We now calculate:
\[\begin{array}{lcll}
t_\sigma(\O\sem{M})& = & 
 s_{\sigma_{i_0}}(\O\sem{N_1}s_{\sigma_1} \ldots
 s_{\sigma_n})\ldots (\O\sem{N_k}s_{\sigma_1}\ldots s_{\sigma_n})\\
& = & \bigwedge_j t_{\tau_j}(\O\sem{N_j}s_{\sigma_1}\ldots s_{\sigma_n})\\
& = & \bigwedge_j t_{\sigma_1 \rightarrow\ldots \rightarrow \sigma_n \rightarrow \tau_j}(\O\sem{N_j}) \quad (\mbox{by the above remark})\\
% && \qquad (\mbox{by the above remark})\\
& = & \top \quad\; (\mbox{by induction hypothesis})
\end{array}\]

Suppose  instead that $M$ is improper. Then it either has the form:
\[\lambda f_1\ldots f_n. \Omega_{\sigma_{i_0}} M_1\ldots M_k\]
or
\[\lambda f_1\ldots f_n. f_{i_0} M_1\ldots M_k\]
where some $N_j \eqdef  \lambda f_1\ldots f_n. M_j$ is improper (and so, by the induction hypothesis $t_{\sigma_1 \rightarrow \ldots \rightarrow \sigma_n \rightarrow \tau_j}(\O\sem{N_j}) = \perp$).

In the first case we have
\[\begin{array}{lcl}
t_\sigma(\O\sem{M}) & = & 
 \O\sem{\Omega_{\sigma_{i_0}}}(\O\sem{N_1}s_{\sigma_1} \ldots s_{\sigma_n})\ldots (\O\sem{N_k}s_{\sigma_1}\ldots s_{\sigma_n})\\
 & = & \perp
\end{array}\]
In the second we have:
\[\begin{array}{lcl}
t_\sigma(\O\sem{M})& = & 
% s_{\sigma_{i_0}}(\O\sem{N_1}s_{\sigma_1} \ldots
% s_{\sigma_n})\ldots (\O\sem{N_k}s_{\sigma_1}\ldots s_{\sigma_n})\\
%& = & \bigwedge_j t_{\tau_j}(\O\sem{N_j}s_{\sigma_1}\ldots s_{\sigma_n})\\
%& = & 
\bigwedge_j t_{\sigma_1 \rightarrow \ldots \rightarrow \sigma_n \rightarrow \tau_j}(\O\sem{N_j})\\
& = & \perp \end{array}\]
\end{proof}
\end{lemma}

\cut{begin{lemma} \label{flowcor} Let  $M,N:\sigma$ be closed $\lambda\Omega^+$-terms such that $\mathcal{O}\sem{M} = \mathcal{O}\sem{N}$. Then $M$ is provably equal in $\lambda\Omega^+$ to a closed $\lambda$-term  if, and only if, $N$ is.
\end{lemma}
\begin{proof} Immediate from Lemma~\ref{flow} and the fact that being provably equal to a closed $\lambda$-term in $\lambda\Omega^+$ is, by Church-Rosser, the same as having a proper normal form.
\end{proof}}
 As a straightforward consequence of Lemma~\ref{flow} we have:
\begin{lemma}  \label{flowcor}  
A closed $\lambda\Omega^+$-term $M\!:\!\sigma$ is provably equal to a closed $\lambda$-term (equivalently, has a 
proper normal form) if, and only if, { $t_\sigma(\O\sem{M}) = \top$}. 
\end{lemma}

\section{Equipotence}% and undefinability}x
 We establish equipotence by using Lemma~\ref{flowcor} to  show the the various translations between our calculi satisfy the conditions of Lemma~\ref{general}. We already know that the identity translation from $\lambda\Omega$ to $\lambda\Omega^+$ does. For the remaining three translations we have:
 \begin{lemma}\label{mainlemma}
 $\;$\\[-1em]
 \begin{enumerate}
 
 \item For any closed $\lambda\Omega^+$-term $M$ and closed $\lambda$-term $N$ of the same type we have:
\[\vdash_{\lambda\Omega^+} M = N \iff \vdash_{\lambda\Omega} \overline{M} = N\]

 \item For  any  closed $\lambda\Omega^+$-term $M$ and closed $\lambda$-term $N$ of the same type we have:
\[\vdash_{\lambda\Omega^+} M = N \iff \vdash_{\lambda\Y} \widehat{M} = N\]
 \item For  any  closed $\lambda\Y$-term $M$ and closed $\lambda$-term $N$ of the same type we have:
\[\vdash_{\lambda\Y} M = N \iff \vdash_{\lambda\Omega^+} \widetilde{M} = N\]
 \end{enumerate}
 \end{lemma}
 \begin{proof} $\;$\\[-1.5em]
 \begin{enumerate}
 %%%%%%%%
 \item We already have the implication from left to right by Lemma~\ref{downsynt}. In the other direction, 
as $\mathcal{O}\sem{M} = \mathcal{O}\sem{\overline{M}}$, and as $\vdash_{\lambda\Omega} \overline{M} = N$, we have 
$t_\sigma(\O\sem{M}) = t_\sigma(\O\sem{N})  = \top$ by Lemma~\ref{flowcor}, and so,  by the same lemma, $\vdash_{\lambda\Omega^+} M = N'$ for some  closed $\lambda$-term $N'$. So then
 $\vdash_{\lambda\Omega} \overline{M} = N'$ by Lemma~\ref{downsynt}.
 But then  $\vdash_{\lambda\Omega}  N' = N$, as $\vdash_{\lambda\Omega} \overline{M} = N$.
 So as  $\vdash_{\lambda\Omega^+} M = N'$ and as $\lambda\Omega^+$ is an extension of $\lambda\Omega$, we have $\vdash_{\lambda\Omega^+} M = N$ as required.
  %%%%%%%%
  \item 
 We already have the implication from left to right by Lemma~\ref{hat}. In the other direction, 
as $\mathcal{O}\sem{M} = \mathcal{O}\sem{\widehat{M}}$, and as $\vdash_{\lambda\Y} \widehat{M} = N$, arguing as in the previous case, employing Lemma~\ref{flowcor}, 
we find that 
$\vdash_{\lambda\Omega^+} M = N'$ for some  closed $\lambda$-term $N'$. So then
 $\vdash_{\lambda\Y} \widehat{M} = N'$ by Lemma~\ref{downsynt}.
 But then  $\vdash_{\lambda\Y}  N' = N$, as $\vdash_{\lambda\Y} \widehat{M} = N$. So, as $\lambda\Y$ is conservative over $\lambda\beta\eta$, and, as  $\vdash_{\lambda\Omega^+} M = N'$, we have $\vdash_{\lambda\Omega^+} M = N$ as required.
  %%%%%%%%
  \item 
  
   We have the implication from right to left by Lemma~\ref{upsynt}. In the other direction 
as $\mathcal{O}\sem{\widetilde{M}} = \mathcal{O}\sem{M}  = \mathcal{O}\sem{N}$, we have 
$\vdash_{\lambda\Omega^+} \widetilde{M} = N'$ for some closed $\lambda$-term $N'$ by Lemma~\ref{flowcor}. 
So then $\vdash_{\lambda\Y} M = N'$ by Lemma~\ref{upsynt}. But then $\vdash_{\lambda\Y }N' = N$,  as $\vdash_{\lambda\Y} M = N$. So, as $\lambda\Y$ is conservative over $\lambda\beta\eta$, and, as  $\vdash_{\lambda\Omega^+} M = N'$, we have $\vdash_{\lambda\Omega^+} M = N$ as required.
 \end{enumerate}
 \end{proof}

With this lemma in  hand, with the identity translation from $\lambda\Omega$ to $\lambda\Omega^+$,  and  with Lemma~\ref{downOmega} on the equipotence of the $\lambda\beta\eta$- and 
$\lambda\Omega$-calculi  for total functions, we  obtain our theorem that recursion does not help:
\begin{theorem} \label{myresult}
Let $\gamma_i: X_i \rightarrow \Lambda_{\sigma_i}\; (i = 1,\ldots,  k)$, and $\gamma: X \rightarrow \Lambda_\sigma$ be coding functions. Then:
%$\,$\\[-1em]
\begin{enumerate}
\item The $\lambda\Y$-calculus is equipotent with the $\lambda\Omega$-calculus for \\$(\gamma_1, \ldots, \gamma_k, \gamma)$-definable  functions.
\item 
%If the type $\sigma_1 \rightarrow \ldots  \rightarrow \sigma_k \rightarrow \sigma$ is inhabited then 
The $\lambda\Y$-calculus is equipotent with the $\lambda\beta\eta$-calculus for\\ $(\gamma_1, \ldots, \gamma_k, \gamma)$-definable total  functions.
\end{enumerate}
\end{theorem}
\newtheorem{fact}{Fact}

We remark that  one might add the converse of the Graph condition, viz:
\[\vdash_{\lambda^+} F\gamma(x_1) \ldots \gamma(x_k) = \gamma(x) \implies f(x_1,\ldots,x_k) = x\]
to the definition of  the $(\gamma_1, \ldots, \gamma_k, \gamma)$-definability of a function $f$ in an extension $\lambda^+$ of the $\lambda\beta\eta$-calculus by a term $F$. With this addition, Lemma~\ref{general} still goes through. Further, the addition makes no difference to definability provided that the coding function $\gamma$ is \emph{$\lambda^+$-injective} by which we mean:
\[\vdash_{\lambda^+} \gamma(x) = \gamma(y) \implies x = y\]
So, with this addition to the definition of definability, Part 1 of Theorem~\ref{myresult} still holds, and Part 2 holds when $\gamma$ is $\lambda\beta\eta$-injective.

Some known undefinabilities for natural number functions follow from Theorem~\ref{myresult}. As we mentioned, Zakrzewski proved that predecessor is not uniformly definable
and Statman proved\footnote{Personal communication, reported in~\cite{FLO83}}, that neither equality nor inequality ($\leq$) are uniformly definable.
%, and
 %Zakrzewski showed that predecessor is not uniformly definable (see~\cite{Zak07}).

These facts follow from the theorem. %corollary. 
For  the three functions are interdefinable via suitable recursions, and so, if one of them was uniformly definable, one could uniformly define arbitrary partial recursive functions in the $\lambda\Y$-calculus.
 They also follow from Statman's result~\cite{Sta04} that it is decidable whether a $\lambda\Y$-term  has a $\lambda\beta\eta$ normal form, since that implies that all definable partial recursive functions have recursive domains.

We can use the fact that arbitrary recursion depth is not available to show that even equality to a fixed number may not be $\sigma$-definable. Define $e_m$ by:
\[e_m(n) = \left \{\begin{array}{ll} 
                            1 & (n = m)\\
                            0 & (\mbox{otherwise})
                         \end{array} \right.\]

\begin{fact} \label{fact1} For any $\sigma$, $e_m$ is not $\sigma$-definable for $m > h(\sigma \rightarrow \sigma)$.
\end{fact}
\begin{proof} We write $\omega_\sigma$ for $\rho_{\Sigma_{\small \mathrm{Nat}},\sigma}$, i.e., for $(\sigma \rightarrow \sigma) \rightarrow (\sigma \rightarrow \sigma)$,  and $\underline{m}_\sigma$ for $\gamma_{\Sigma_{{\small \mathrm{Nat}}},\sigma}(m)$, i.e.,  for $\lambda f^{\sigma\rightarrow \sigma}.\lambda x^\sigma.f^m(x)$. Suppose, for the sake of contradiction, that a term $E_m$  $\sigma$-defines $e_m$ for 
$m > h(\sigma \rightarrow \sigma)$. Let $Z = \Y_{\omega_\sigma\rightarrow \omega_\sigma}F$ where 
\[F =  \lambda f x. \mathtt{if} \, E_m x \,\mathtt{then}\, \underline{0}_\sigma \,\mathtt{else} \, f(x + \underline{1}_\sigma)\]
(Recall that the conditional %, treating $1$ and $0$ as true and false,
 is an extended polynomial.)

We have $\vdash_{\lambda\Y} Z\underline{0}_\sigma = \underline{0}_\sigma$. However, using induction on $k$, one shows that
 $\vdash_{\lambda\Omega^+}\widetilde{\Y}_\sigma^{(k)} F\underline{l}_\sigma = \Omega_\sigma$ when $k+l < m$.
So $\vdash_{\lambda\Omega^+}\widetilde{\Y}_\sigma^{(h(\sigma \rightarrow \sigma))}F\underline{0}_\sigma = \Omega_\sigma$ in particular. Hence, by Lemma~\ref{upsynt}, we have 
$\vdash_{\lambda\Y}\Y_\sigma F\underline{0}_\sigma = \Omega_\sigma$. So $\vdash_{\lambda\Y} \underline{0}_\sigma = \Omega_{\sigma}$ contradicting the consistency of $\lambda\Y$.
\end{proof}

It follows that comparisons to a suitably large fixed number are not $\sigma$-definable either. Using a similar bounded recursion depth argument one can obtain  Zakrzewski's result~\cite{Zak07}  that $\lfloor{n/2}\rfloor$ is not uniformly definable.

Using the flow analysis technique we can also prove Statman's decidability result. We have a version of Lemma~\ref{flowcor} for the $\lambda\Y$-calculus:
\begin{lemma} \label{statmanresult}
 A closed $\lambda\Y$-term $M\!:\!\sigma$ is provably equal to a closed $\lambda$-term (equivalently, has a 
 $\lambda\beta\eta$ normal form) if, and only if, $t_\sigma(\O\sem{M}) = \top$. 
\end{lemma}
\begin{proof} The implication from left to right follows from Lemma~\ref{flow}. In the other direction, as 
$\O\sem{M} = \O\sem{\widetilde{M}}$ and $\widetilde{M}$ has a normal form, the same lemma tells us that $\widetilde{M}$ has a proper normal form. The conclusion follows, by Lemma~\ref{upsynt}.
\end{proof}
Statman's result then follows as it is evidently decidable for closed $\lambda\Y$-terms $M$ whether or not $t_\sigma(\O\sem{M}) = \top$.
We remark that, using a different flow analysis, one can obtain another of Statman's results, that it is decidable whether a given $\lambda\Y$-term  has a head normal form. One takes:
\[t_\sigma(f) = fs_{\sigma_1}\ldots s_{\sigma_n}
\qquad  \qquad s_\sigma f_1\ldots f_n = \top\]
\section*{Acknowledgements}

This paper is based on a manuscript I wrote  on a visit to MIT in 1982, a visit for which I am grateful to Albert Meyer. However, the manuscript sank below the waves until Pawe{\l} Urzyczyn very kindly sent me a copy. I am delighted to thereby have had the opportunity to have something to tell Jonathan.

The paper will appear in \emph{The Mathematical Foundation of Computation}, a Festschrift for Jonathan Seldin on the occasion of his 80th birthday, edited by Fairouz Kamareddine. I am grateful to College Publications for permission to post it on the web, and I am grateful to Pawel for many helpful comments on drafts.

\end{document}